\documentclass[aps,prc,twocolumn,superscriptaddress,showpacs]{revtex4}
\usepackage{graphicx}

\begin{document}


\title{Lifetime of $^{19}$Ne$^\ast$(4.03 MeV)}


\author{R. Kanungo}
\affiliation{TRIUMF, 4004 Wesbrook Mall, Vancouver BC V6T 2A3 Canada}
\author{T. K. Alexander}
\affiliation{Deep River, ON, Canada}
\author{A. N. Andreyev}
\affiliation{TRIUMF, 4004 Wesbrook Mall, Vancouver BC V6T 2A3 Canada}
\affiliation{ Department of Chemistry, Simon Fraser University, Burnaby BC, Canada}
\author{G. C. Ball}
\affiliation{TRIUMF, 4004 Wesbrook Mall, Vancouver BC V6T 2A3 Canada}
\author{R. S. Chakrawarthy}
\affiliation{TRIUMF, 4004 Wesbrook Mall, Vancouver BC V6T 2A3 Canada}
\affiliation{ Department of Chemistry, Simon Fraser University, Burnaby BC, Canada}
\author{M. Chicoine}
\affiliation{D\'{e}partement de Physique, Universit\'{e} de Montr\'{e}al, Montr\'{e}al QB, Canada}
\author{R. Churchman}
\affiliation{TRIUMF, 4004 Wesbrook Mall, Vancouver BC V6T 2A3 Canada}
\author{B. Davids}
\affiliation{TRIUMF, 4004 Wesbrook Mall, Vancouver BC V6T 2A3 Canada}
\author{J. S. Forster}
\affiliation{D\'{e}partement de Physique, Universit\'{e} de Montr\'{e}al, Montr\'{e}al QB, Canada}
\author{S. Gujrathi}
\affiliation{D\'{e}partement de Physique, Universit\'{e} de Montr\'{e}al, Montr\'{e}al QB, Canada}
\author{G. Hackman}
\affiliation{TRIUMF, 4004 Wesbrook Mall, Vancouver BC V6T 2A3 Canada}
\author{D. Howell}
\affiliation{TRIUMF, 4004 Wesbrook Mall, Vancouver BC V6T 2A3 Canada}
\affiliation{Department of Physics, Simon Fraser University, Burnaby BC, Canada}
\author{J. R. Leslie}
\affiliation{Department of Physics, Queen's University, Kingston ON, Canada}
\author{A. C. Morton}
\affiliation{TRIUMF, 4004 Wesbrook Mall, Vancouver BC V6T 2A3 Canada}
\author{S. Mythili}
\affiliation{TRIUMF, 4004 Wesbrook Mall, Vancouver BC V6T 2A3 Canada}
\affiliation{Department of Physics and Astronomy, University of British Columbia, Vancouver BC, Canada}
\author{C. J. Pearson}
\affiliation{TRIUMF, 4004 Wesbrook Mall, Vancouver BC V6T 2A3 Canada}
\author{J. J. Ressler}
\affiliation{ Department of Chemistry, Simon Fraser University, Burnaby BC, Canada}
\author{C. Ruiz}
\affiliation{TRIUMF, 4004 Wesbrook Mall, Vancouver BC V6T 2A3 Canada}
\affiliation{ Department of Chemistry, Simon Fraser University, Burnaby BC, Canada}
\author{H.~Savajols}
\affiliation{TRIUMF, 4004 Wesbrook Mall, Vancouver BC V6T 2A3 Canada}
\affiliation{GANIL, Caen, France}
\author{M. A. Schumaker}
\affiliation{Physics Department, University of Guelph, Guelph ON, Canada}
\author{I. Tanihata}
\author{P. Walden}
\author{S. Yen}
\affiliation{TRIUMF, 4004 Wesbrook Mall, Vancouver BC V6T 2A3 Canada}

\date{\today}

\begin{abstract}
The Doppler-shift attenuation method was applied to measure the lifetime of the 4.03 MeV state in $^{19}$Ne. Utilizing a $^3$He-implanted Au foil as a target, the state was populated using the $^{20}$Ne($^3$He,$\alpha)^{19}$Ne reaction in inverse kinematics at a $^{20}$Ne beam energy of 34 MeV. De-excitation $\gamma$ rays were detected in coincidence with $\alpha$ particles. At the 1$\sigma$ level, the lifetime was determined to be 11$^{+4}_{-3}$~fs and at the 95.45\% confidence level the lifetime is 11$^{+8}_{-7}$~fs.
\end{abstract}


\maketitle

\section{Introduction}
The $^{15}$O($\alpha,\gamma)^{19}$Ne reaction leads to the initial breakout from the hot CNO cycles that operate in Type I x-ray bursts, which are thermonuclear runaways on accreting neutron stars in binary star systems \cite{wiescher99}. Recent calculations have suggested that if the rate of this reaction were below a certain threshold, the periodic x-ray bursts observed from some accreting neutron stars would not occur \cite{fisker04}. Hence the rate of this reaction is of considerable importance. However, direct measurements at the relevant energies would require (radioactive) $^{15}$O beams of high intensity not presently available. Since the first theoretical investigation of this reaction \cite{wagoner69}, experimental data on the radiative and $\alpha$ widths of excited states in $^{19}$Ne have been used to better constrain its rate. As was pointed out first in Ref.\ \cite{langanke86}, at temperatures below 0.6 GK the reaction rate is dominated by resonant capture to the first state above the $\alpha$-emission threshold, lying at an excitation energy of 4.03 MeV.

The decay widths of the 4.03 MeV state in $^{19}$Ne have until recently remained elusive. Its $\alpha$ width, $\Gamma_\alpha$, has been experimentally unobservable on account of its small value. All published attempts to measure the $\alpha$-decay branching ratio B$_\alpha \equiv \Gamma_\alpha/\Gamma$ have yielded only upper limits \cite{magnus90, laird02, davids03, rehm03,davids03a,visser04}. Therefore the reduced $\alpha$ width of the analog state at 3.91 MeV in $^{19}$F \cite{mao95,mao96} has been used to estimate the $\alpha$ width. Early theoretical estimates of the reaction rate assumed the radiative widths of the $^{19}$Ne and $^{19}$F analog states to be equal \cite{wallace81,langanke86}. Despite attempts to measure the radiative width of the state in $^{19}$Ne itself that resulted in lower and upper limits \cite{davidson73,hackman00}, the analog state has been the most reliable source of experimental information on the radiative width. With the recent report of the first measurement of the lifetime of the 4.03 MeV state, as well as more precise determinations of the excitation energies of this and other states in $^{19}$Ne \cite{tan05}, the experimental situation has improved dramatically. We report here a second successful measurement of the lifetime of the 4.03 MeV state in $^{19}$Ne, using a different reaction in which the recoil velocity was higher, allowing for a more precise lifetime determination.

\section{Experiment and Analysis}

\begin{figure}
\includegraphics[width=\linewidth]{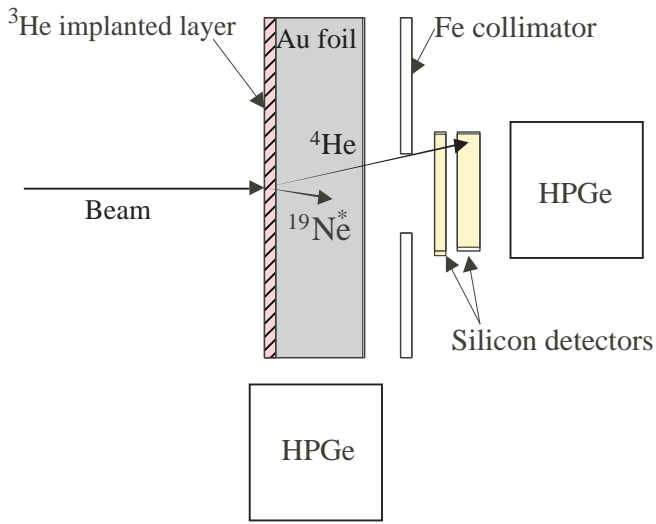}
\caption{(Color online) A schematic view of the experiment. The dimensions are not drawn to scale.}
\label{fig1}
\end{figure}

The experiment was performed at the TRIUMF-ISAC facility using the Doppler-shift attenuation method. A 34 MeV $^{20}$Ne beam was incident on a $^3$He-implanted Au foil target, populating the level of interest via the $^3$He($^{20}$Ne,$\alpha)^{19}$Ne reaction. The $^{20}$Ne beam and recoiling $^{19}$Ne nuclei were stopped in the Au foil. The average beam intensity was 10 particle nA. A schematic depiction of the experimental setup is shown in Fig.\ \ref{fig1}. De-excitation $\gamma$ rays were detected in coincidence with $\alpha$ particle ejectiles using an 80\% high-purity germanium (HPGe) coaxial detector placed at 0$^\circ$ with respect to the beam axis. The lifetime was determined from a lineshape analysis of this $\gamma$-ray energy spectrum. A second HPGe detector placed at  90$^\circ$  was used as a reference detector to measure the unshifted peak energies. The detectors were located 9 cm from the target. The energy calibration of the HPGe detectors was performed using a $^{56}$Co source whose highest energy $\gamma$ ray is 3.2 MeV. This calibration was extrapolated linearly to higher energies. The energies of $\gamma$ rays from the source were measured before and after the experiment and were found to differ by less than 1 keV.

The scattering chamber was designed with a cold trap to ensure a clean target surface and also to prevent losses of the implanted $^3$He. This was achieved using a narrow differential pumping aperture followed by a copper cylinder enclosing the path of the beam to the target. The copper cylinder was cooled using liquid nitrogen. To avoid any condensation of impurities on the surface of the target, the copper cylinder was not in direct contact with the target ladder. Indirect contact of the cold copper cylinder with the copper target ladder was achieved using BeCu fingers mounted on a boron nitride plate, which provided electrical isolation as well. This arrangement maintained a temperature difference between the copper cylinder and the target ladder. In this way the target was cooled below room temperature to ensure that $^3$He did not diffuse out when heated by bombardment with a beam power of up to 0.3 W. Moreover, the colder surfaces surrounding the target foil and the beam path in front of it reduced the buildup of carbon and other contaminants on the target itself during the experiment. 

The target foil was prepared at the Universit\'{e} de Montr\'{e}al by implanting 30 keV $^3$He ions into a 12.5 $\mu$m thick Au foil, yielding an areal number density of 6$\times$10$^{17}$~cm$^{-2}$. The implantation region was 0.1 $\mu$m thick. Similar implanted foils prepared at the Chalk River laboratories were used in earlier femtosecond lifetime measurements \cite{keinonen81}. The foil was found to contain some surface deposits of carbon after the implantation process. The concentration of $^3$He in the foil was monitored via yields of elastically scattered $^3$He and was found to remain unchanged during the experiment. The beam was collimated to a 2 mm diameter spot on the target.

\begin{figure}
\includegraphics[width=\linewidth]{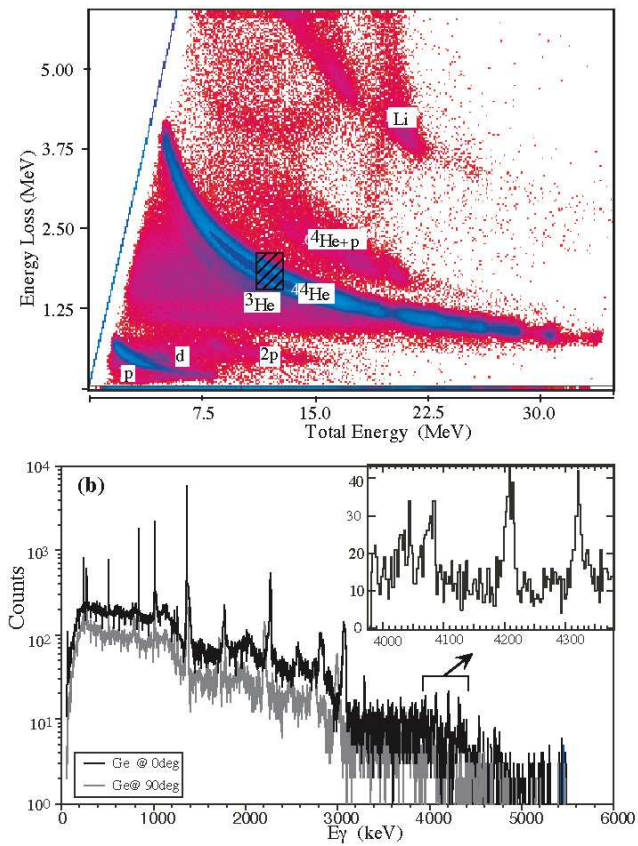}
\caption{(Color online) (a) The particle identification spectrum in the silicon detectors in coincidence with either of the HPGe detectors. (b) The $\gamma$-ray spectra in the HPGe detectors in coincidence with  $\alpha$ particles from the hatched area of part (a). The inset shows the energy range around 4 MeV in the 0$^\circ$ HPGe detector}
\label{fig2}
\end{figure}

The $\alpha$ particles were identified by means of the energy loss ($\Delta$E) and total energy (E) correlation using a silicon detector telescope. The telescope consisted of a 25 $\mu$m thick silicon detector for $\Delta$E measurement and a 500$\mu$m silicon detector for E measurement. Both the detectors were standard, circular, ORTEC transmission-type detectors with an active area of 150 mm$^2$. The detector telescope subtended a solid angle of 360 msr and was centered about the beam axis, allowing the detection of $\alpha$ particles with scattering angles less than 20$^\circ$. The $\Delta$E-E particle identification spectrum obtained in coincidence with at least one HPGe detector is shown in Fig.\ \ref{fig2}(a). A wide range of $\alpha$ particle energies arose from fusion-evaporation reactions with the carbon contaminant on the foil. The region in which $\alpha$ particles from the 4.03 MeV level in $^{19}$Ne can be found is marked by the hatched band. The $\gamma$-ray energy spectrum obtained from the HPGe detectors in coincidence with $\alpha$ particles falling within the hatched energy band is shown in Fig.\ \ref{fig2}(b). Several peaks arising from fusion evaporation products can be identified. The inset shows the E$_\gamma \sim $ 4 MeV region from the 0$^\circ$ HPGe detector, revealing several distinct peaks. The two peaks observed between 4.0 and 4.1 MeV in the 0$^\circ$ detector are inconsistent with known $\gamma$ rays from $^{19}$Ne, while those at 4.2 and 4.3 MeV are consistent with known $^{19}$Ne transitions Doppler shifted appropriately for recoils moving with $v = 0.04 c$.

\begin{figure}
\includegraphics[width=\linewidth]{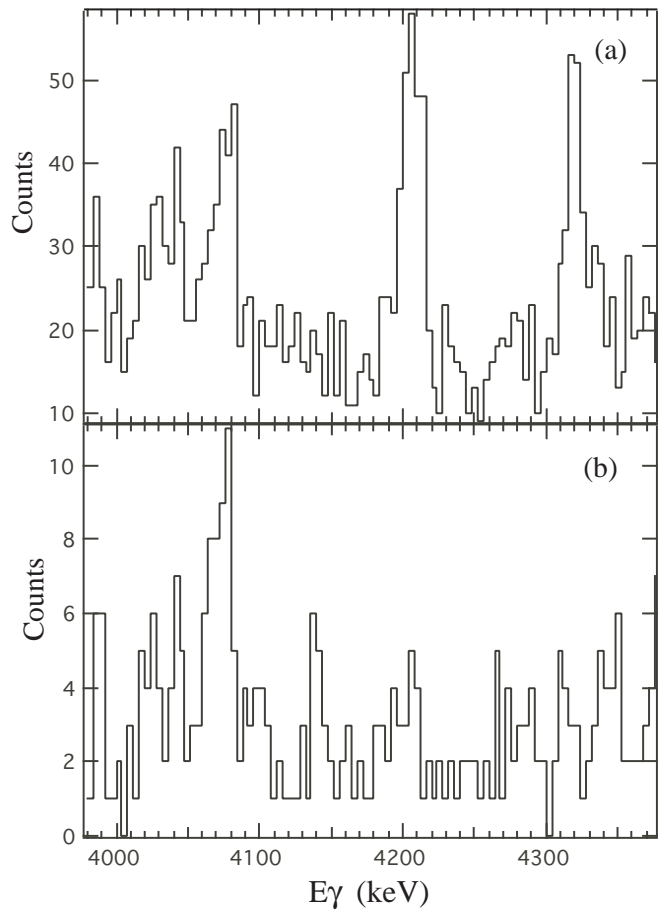}
\caption{Gamma ray energy spectra from the 0$^\circ$ HPGe detector obtained in coincidence with $\alpha$ particles with total energies between (a) 11 and 13 MeV, and (b) 13 and 14 MeV.}
\label{fig3}
\end{figure}

The $\gamma$ ray energy spectrum from the 0$^\circ$ HPGe detector measured in coincidence with $\alpha$ particles in two different total $\alpha$ energy (E$_\alpha$) ranges is shown in Fig.\ \ref{fig3}. The $\alpha$ particles corresponding to 4.03 MeV excitations in $^{19}$Ne falling within the angular acceptance of our setup have total energies between 10 and 13.5 MeV. Examining the $\gamma$ spectrum with a gate on E$_\alpha$ between 11 and 13 MeV (a slightly smaller range than that expected in order to improve the signal/noise ratio), shown in Fig.\ \ref{fig3}(a), we see two peaks corresponding to Doppler-shifted $\gamma$ rays from $^{19}$Ne. The peak observed at 4.2 MeV in the 0$^\circ$ HPGe detector corresponds to the Doppler-shifted direct $\gamma$ transition from the 4.03 MeV level to the ground state of $^{19}$Ne, and the peak at 4.32 MeV is consistent with the Doppler-shifted 4.14 MeV de-excitation $\gamma$ ray from the 4.38 MeV level to the 238 keV level in $^{19}$Ne. Gating on $\alpha$ particles from 13-14 MeV, shown in Fig.\ \ref{fig3}(b), we see little evidence for Doppler-shifted $\gamma$ rays from the 4.03 MeV and 4.38 MeV states, just as we would expect.

\begin{figure}
\includegraphics[width=\linewidth]{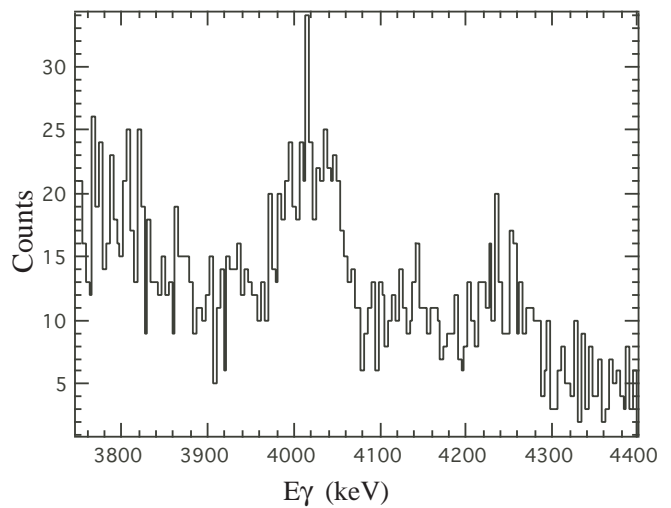}
\caption{The $\gamma$ ray energy spectrum from the 90$^\circ$ HPGe detector measured in coincidence with $\alpha$ particles having total energies between 11 and 13 MeV. The Doppler-broadened peak around 4.03 MeV is consistent with the direct transition in $^{19}$Ne from the 4.03 MeV level to the ground state.}
\label{fig4}
\end{figure}

Figure \ref{fig4} shows the energy spectrum from the 90$^\circ$ HPGe detector in coincidence with $\alpha$ particles having energies between 11 and 13 MeV. The spectrum shows a peak at 4.03 MeV, consistent with the direct transition from the 4.03 MeV state to the ground state, which is Doppler-broadened because of the large angular acceptance of the HPGe detector. The absence of a sharp peak at 4.2 MeV indicates that there are no $\gamma$ rays from a long-lived, stopped contaminant in the region of interest.

\begin{figure}
\includegraphics[width=\linewidth]{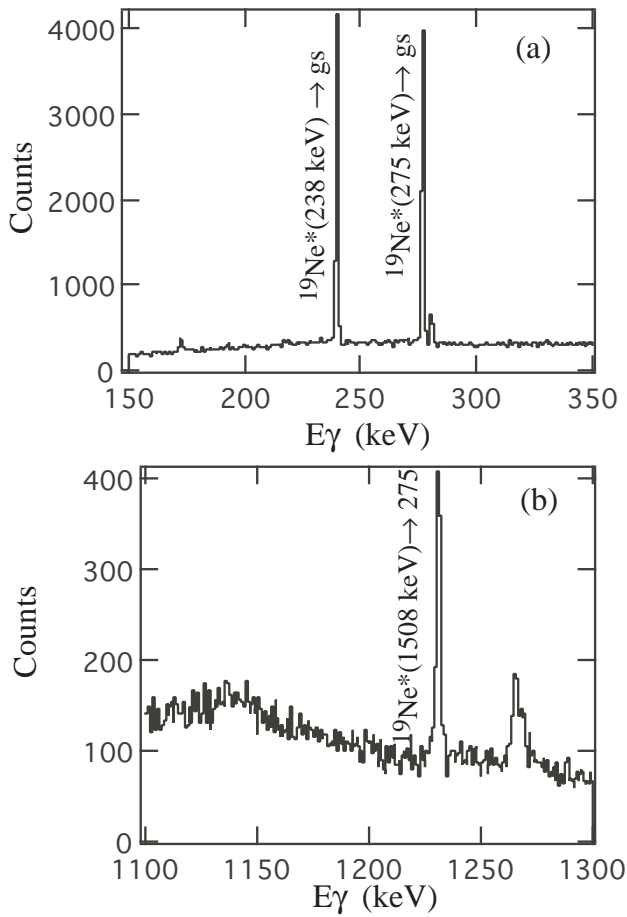}
\caption{$\gamma$ ray spectra from the 0$^\circ$ HPGe detector obtained in coincidence with $\alpha$ particles. Part (a) shows the 150-350 keV region and (b) the 1100-1300 keV region.}
\label{fig5} 
\end{figure}

\begin{figure}
\includegraphics[width=\linewidth]{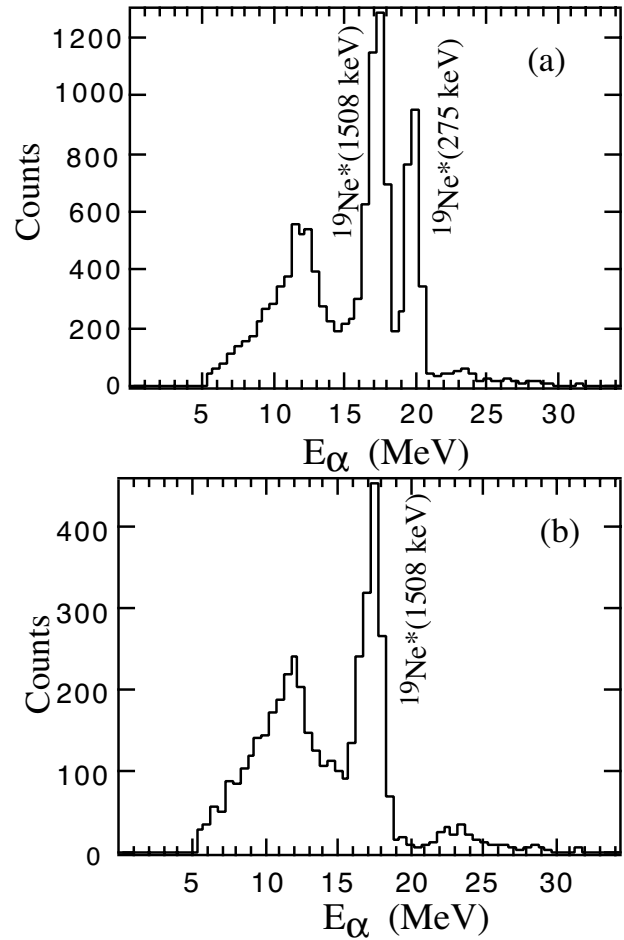}
\caption{Total energy spectra of $\alpha$ particles observed in coincidence with the $\gamma$-ray peaks at (a) 275 keV and (b) 1233 keV.}
\label{fig6} 
\end{figure}

\begin{figure}
\includegraphics[width=\linewidth]{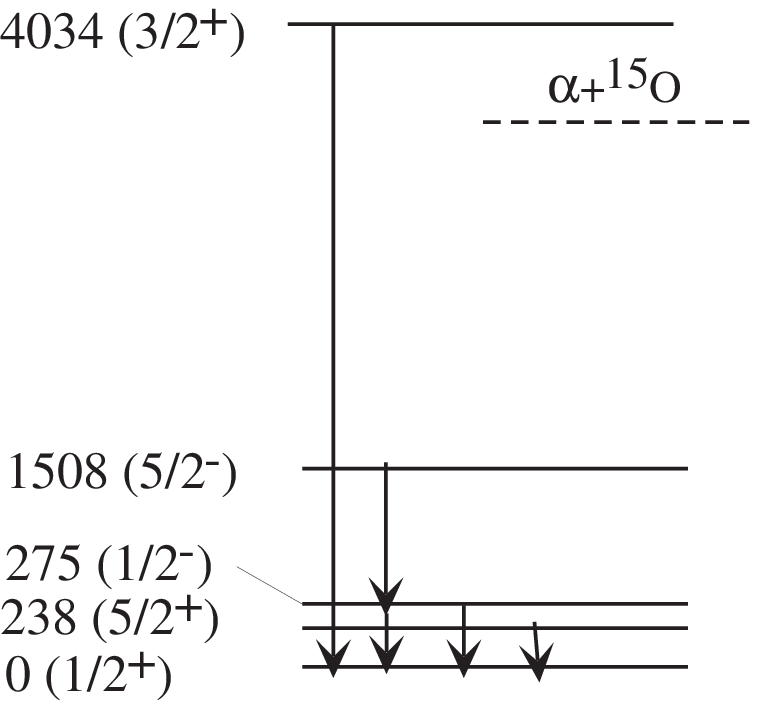}
\caption{Schematic rendering of some of the $^{19}$Ne energy levels and $\gamma$-ray transitions observed in the experiment. Energies are given in keV.}
\label{levels} 
\end{figure}

Looking at the low energy range of the $\gamma$-ray spectrum from the 0$^\circ$ HPGe detector obtained in coincidence with all $\alpha$ particles, shown in Fig.\ \ref{fig5}(a) and (b), clear signatures for population of the first three excited states in $^{19}$Ne are also apparent. Gamma rays from the first excited state appear at 238 keV. The 275 keV $\gamma$ ray visible in Fig.\ \ref{fig5}(a) is from the de-excitation of the second excited state in $^{19}$Ne. Higher-lying states in $^{19}$Ne also decay through these two levels with finite decay branches. One such decay branch is from the 1508 keV state in $^{19}$Ne which decays by the emission of a 1233 $\gamma$ ray to the 275 keV level, as can be seen in Fig.\ \ref{fig5}(b). This branch can be confirmed by looking at the E$_\alpha$ spectrum gated on the 275 and 1233 keV $\gamma$ rays, depicted in Fig.\ \ref{fig6}(a) and (b), respectively. The $\alpha$ particle energy spectrum shows a peak at 17 MeV when gated on either the 275 or 1233 keV $\gamma$ rays, which corresponds to the population of the third excited state in $^{19}$Ne, lying at 1508 keV. Additionally the E$_\alpha$ peak  at 20 MeV due to the second excited state in $^{19}$Ne is seen to be clearly separated from the peak corresponding to the third excited state. For clarity, a level diagram indicating the $^{19}$Ne states and transitions discussed in the paper is shown in Figure \ref{levels}.

The response function of the experimental setup includes the effects of kinematic broadening arising from the finite angular acceptance of the silicon detectors as well as Doppler broadening due to the finite opening angle of the HPGe detector. The intrinsic lineshape of the HPGe detector, measured using a 3.2 MeV $\gamma$ ray from the de-excitation of a $^{56}$Fe level populated in $^{56}$Co electron capture, was included in the response function. The line width due to the intrinsic resolution (3.5 keV FWHM at 4 MeV) is much narrower than the Doppler broadening observed in the measurement. The detection efficiency as a function of the emission angle of a 4 MeV $\gamma$ ray for the angular range subtended by the 0$^\circ$ HPGe detector was taken into account by a GEANT4 simulation \cite{geant4} appropriate for the geometry of the setup. The effects of different $\gamma$-ray angular distributions were investigated and found to have a negligible influence on the calculated lineshape. We estimate the 1$\sigma$ uncertainty in the relative detection efficiency to be $\pm$ 5\%, which is the result of uncertainties in the geometry of the setup.

Experimental data exist on the stopping powers of heavy ions in Au at this recoil energy. These data have been used to constrain theoretical stopping power calculations. We used the parametrization of Ziegler \cite{ziegler04}. This result was compared to calculations based on the measurements of the Chalk River group \cite{forster76}. The difference between the two stopping powers is less than 10\%; based on this we estimate a 1$\sigma$ stopping power uncertainty of $\pm$ 5\%. The analysis also took into account the change in stopping power in the $^3$He-implanted region of the Au foil following the prescription outlined in Ref.\ \cite{alexander81}.   

\begin{figure}
\includegraphics[width=\linewidth]{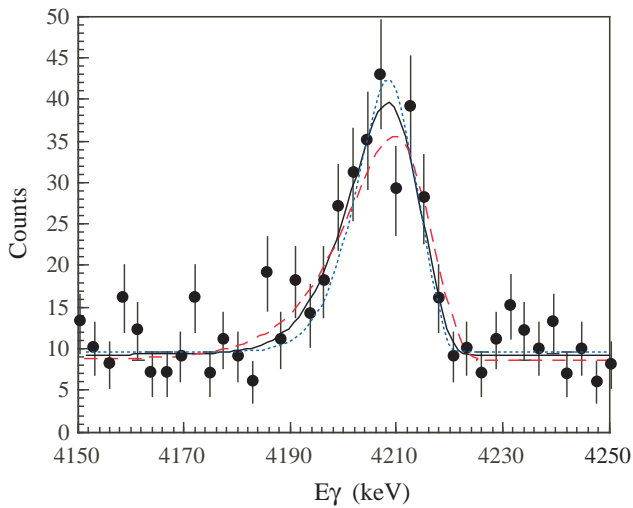}
\caption{(Color online) The lineshape data observed in the 0$^\circ$ HPGe detector for the Doppler-shifted 4.03 MeV $\rightarrow$ ground state $\gamma$-ray transition in $^{19}$Ne. The curves show lineshapes for the best fit (solid line, $\tau$ = 11 fs), the 2$\sigma$ lower limit (dotted line, $\tau$ = 4 fs), and the 2$\sigma$ upper limit (dashed line, $\tau$ = 19 fs) lifetimes determined from a $\chi^2$ minimization.}
\label{lineshape} 
\end{figure}

The lifetime was determined from a $\chi^2$ minimization using the lineshape calculated with
the computer program described in Ref.\ \cite{forster79}, taking into account the initial velocity distribution of the $^{19}$Ne recoils, the intrinsic lineshape of the HPGe detector, and the swelling of the $^3$He-implanted target layer. Apart from the lifetime, the free parameters in the $\chi^2$ search were the overall normalization, the flat background level and the centroid position. We allowed the centroid position to be a free parameter in order to account for any possible shift which could arise due to energy calibration uncertainties.

The $\chi^2$ minimization led to a lifetime for the 4.03 MeV level in $^{19}$Ne of $\tau = 11^{+4}_{-3}$~fs at the 1$\sigma$ level; this uncertainty is the result of several contributions, the most important of which are stopping power ($< \pm$ 1~fs), relative detection efficiency ($\pm$ 1~fs), statistics ($\pm$ 2~fs), and the peak centroid position, which dominates the remaining uncertainty. At the 2$\sigma$ level, $\tau = 11^{+8}_{-7}$~fs. The lineshapes corresponding to the best fit and the 2$\sigma$ upper and lower limit lifetimes are shown in Fig.\ \ref{lineshape}. This value is in excellent agreement with the first lifetime measurement \cite{tan05}, which yielded 13$^{+9}_{-6}$ fs (1$\sigma$). The higher precision of the present result arises from the fact that the $^{19}$Ne recoil velocity and the corresponding Doppler shift in this experiment were much larger than those in the measurement of Ref.\ \cite{tan05}, which used the $^{17}$O($^3$He,$n)^{19}$Ne reaction at 3 MeV. The decay in flight produced a clearly asymmetric lineshape, with a long low-energy tail compared with the more sharply rising edge on the high energy side of the peak in Fig.\ \ref{lineshape}. Reducing the angular acceptance of the detectors would improve the precision of the lifetime determination at the cost of statistics, but this is impractical given the low yield observed in the present experiment. Efforts are underway to reduce the experimental uncertainty by using cleaner target foils to reduce background and finding the optimum beam energy for the reaction to increase yield.

\section{Summary and Conclusions}

The lifetime of the 4.03 MeV state in $^{19}$Ne was measured via the Doppler shift attenuation method. Populating the state using the $^3$He($^{20}$Ne,$\alpha)^{19}$Ne reaction at a beam energy of 34 MeV, we stopped the recoils in the $^3$He-implanted Au target foil and detected de-excitation $\gamma$ rays in coincidence with $\alpha$ particles. The substantial Doppler shift of the recoils allowed a relatively precise determination of the lifetime. At the 1$\sigma$ level, the lifetime was determined to be 11$^{+4}_{-3}$ fs; at the 95.45\% confidence level $\tau$~=~11$^{+8}_{-7}$~fs. The lifetime reported here agrees well with both the measurement of Ref.\ \cite{tan05} and the lifetime of the $^{19}$F analog, $\tau$~=~9(5)~fs \cite{tilley95}, further bolstering the evidence that isospin is a good symmetry in the T~=~1/2, A~=~19 system \cite{davids03a}.

\begin{figure}
\includegraphics[width=\linewidth]{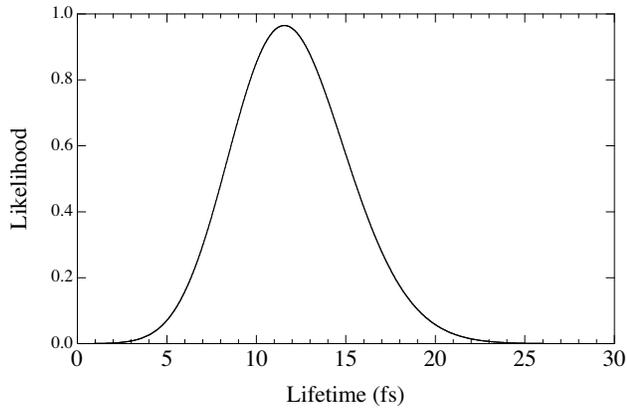}
\caption{The joint likelihood for the lifetime of the 4.03 MeV state in $^{19}$Ne, taking into account the present measurement and that of Tan \emph{et al.} \protect{\cite{tan05}}.}
\label{joint} 
\end{figure}

By combining the present lifetime determination with that of Ref.\ \cite{tan05}, we can tighten the experimental constraints on the radiative width of the 4.03 MeV state. Using the $\chi^2$ information given in the Tan \emph{et al.} paper \cite{tan05} and that from the present measurement, it is possible to construct the joint likelihood for the lifetime, taking into account the data of both experiments. This joint likelihood is shown in Fig.\ \ref{joint} and peaks around 12 fs. When the two experiments are combined, the lifetime is constrained to lie within 3 and 22 fs at the 99.73\% confidence level.

Despite the fact that the $\alpha$-decay branching ratio B$_\alpha$ is presently constrained experimentally only by an upper limit, the two reported measurements of the lifetime of the 4.03 MeV state in $^{19}$Ne and the upper limit on B$_\alpha$ reported in Ref.\ \cite{davids03a} allow us to place an experimental upper limit on $\Gamma_\alpha = B_\alpha \hbar / \tau$. Using the 3$\sigma$ upper limit on B$_\alpha$ and the 3$\sigma$ lower limit on $\tau$, at the 99.73\% confidence level $\Gamma_\alpha < 200$ $\mu$eV. This implies a 3$\sigma$ upper limit on the resonance strength, and thereby the $^{15}$O($\alpha,\gamma)^{19}$Ne reaction rate at T $< 0.6$ GK, approximately 3 times smaller than the 3$\sigma$ upper limit quoted in Ref.\ \cite{davids03a}. Since the experimental upper limit on the rate has decreased, the conclusion reached in that work that this reaction probably plays no significant role in classical novae remains valid, consistent with the conclusions drawn in Refs.\ \cite{wiescher99,iliadis02}. Measurements of the $\alpha$-decay branching ratios of the 4.03 and 4.38 MeV states will be required to more precisely identify the importance of the $^{15}$O($\alpha,\gamma)^{19}$Ne reaction in x-ray bursts.

\begin{acknowledgments}
This work was generously supported by the Natural Sciences and Engineering Research Council of Canada. TRIUMF receives federal funding via a contribution agreement through the National Research Council of Canada. RK is grateful to A.~C.~Shotter for helpful discussions and BD would like to thank R.~H.~Cyburt for enlightening discussions on statistics.
\end{acknowledgments}

\bibliography{19ne}
\end{document}